\documentclass[pdflatex,sn-mathphys-num]{sn-jnl}

\usepackage[T1]{fontenc} 

\usepackage{graphicx}%
\usepackage{multirow}%
\usepackage{amsmath,amssymb,amsfonts}%
\usepackage{amsthm}%
\usepackage{mathrsfs}%
\usepackage[title]{appendix}%
\usepackage{xcolor}%
\usepackage{textcomp}%
\usepackage{manyfoot}%
\usepackage{booktabs}%
\usepackage{algorithm}%
\usepackage{algorithmicx}%
\usepackage{algpseudocode}%
\usepackage{listings}%

\theoremstyle{thmstyleone}%
\theoremstyle{thmstyletwo}%

\theoremstyle{thmstylethree}%

\newgeometry{margin=2.5cm}

\begin{document}

\title[Transversal Gates and Magic State Distillation in an Optimally Synthesized Spin-Qubit Shuttling Bus]
{Transversal Gates and Magic State Distillation in an Optimally Synthesized Spin-Qubit Shuttling Bus}

\author*[1]{\fnm{Pau} \sur{Escofet}}\email{pau.escofet@upc.edu}
\author[2]{\fnm{Andrii} \sur{Semenov}}
\author[2]{\fnm{Niall} \sur{Murphy}}
\author[2]{\fnm{Elena} \sur{Blokhina}}
\author[3]{\fnm{Carmen G.} \sur{Almudéver}}
\author[1]{\fnm{Sergi} \sur{Abadal}}
\author[1]{\fnm{Eduard} \sur{Alarcón}}

\affil[1]{\orgname{Universitat Politècnica de Catalunya}, \orgaddress{\city{Barcelona}, \country{Spain}}}
\affil[2]{\orgname{Equal1 Labs}, \orgaddress{\city{Dublin}, \country{Ireland}}}
\affil[3]{\orgname{Universitat Politècnica de València}, \orgaddress{\city{Valencia}, \country{Spain}}}


\abstract{Fault-tolerant quantum computing requires not only reliable logical qubit storage, but also the ability to perform high-fidelity logical operations between error-corrected qubits at scale. While much of the existing literature focuses on optimizing syndrome extraction for a single logical qubit, the co-design of physical architectures that support both robust error correction and efficient logical computation remains an open challenge. In this work, we propose a multi-qubit spin-qubit shuttling bus architecture that addresses both requirements simultaneously. The architecture optimizes the physical qubit layout for syndrome extraction and supports transversal two-qubit logical gates between an arbitrary number of logical qubits, achieving all-to-all logical connectivity through coherent spin shuttling. We further propose an ancilla-sharing scheme that encodes multiple logical qubits within a single logical element, compressing the physical footprint of the processor and improving long-range gate fidelity. Extending the architecture from a one-dimensional bus to a two-dimensional grid of shuttling tracks reduces the inter-qubit distance, yielding consistent improvements in logical error. Finally, we apply the Quantum Reverse Mapping methodology at the logical level to optimize the layout of a \textit{15-to-1} magic state distillation circuit, demonstrating how the transversal capabilities of the proposed architecture can be leveraged for universal fault-tolerant computation. Taken together, these results establish a principled co-design framework that bridges the physical, error-correction, and logical computation layers of the quantum stack, and demonstrate that spin-qubit shuttling architectures are a viable and flexible substrate for scalable fault-tolerant quantum computation.}



\maketitle


\section{Introduction}
\label{sec:intro}

Quantum computing architectures will only scale to the requirements needed for fault-tolerant computational advantage through the use of quantum error correction (QEC)~\cite{preskill2018quantum, arute2019quantum, eisert2025mind}. The core idea behind QEC is to encode the state of one logical qubit redundantly across many physical qubits, such that errors affecting individual physical qubits can be detected and corrected without disturbing the encoded logical information \cite{nielsen2010quantum, roffe2019quantum}. This is achieved by repeatedly measuring a set of multi-qubit operators (the stabilizers) whose outcomes, called the syndrome, reveal the location and type of errors without collapsing the logical state. A classical decoder processes the syndrome and determines the appropriate correction to apply, allowing the logical qubit to be maintained with an error rate that decreases exponentially with the number of physical qubits, provided the physical error rate lies below a code-specific threshold.

Several families of QEC codes have been proposed and studied. Topological codes, including the surface code~\cite{bravyi1998quantum, dennis2002topological, fowler2009high, fowler2012surface} and color codes~\cite{bombin2006topological, fowler2011two}, stand out for their high error-correction thresholds, their reliance on local nearest-neighbor interactions, and the availability of efficient and scalable decoding algorithms~\cite{higgott2022pymatching, higgott2025sparse}. Recently, the quantum low-density parity-check (qLDPC) family of codes~\cite{breuckmann2021quantum, panteleev2022asymptotically} has gained significant attention, which promise to be asymptotically "good" codes, offering a substantially lower physical-to-logical qubit overhead than topological codes, at the cost of more complex, non-local stabilizer measurements.

The choice of QEC code, and how it is integrated with the physical hardware, has deep implications across many layers of the computing stack, from how syndrome extraction is obtained, to which logical operations are natively available and how they can be efficiently executed. These implications have motivated a growing body of work on the co-design of QEC codes and physical architectures~\cite{thantharate2023q, lin2024codesign, stein2025hetec, yoder2025tour, webster2026pinnacle}, in which the hardware layout and the code structure are jointly optimized, typically targeting either syndrome extraction efficiency or the performance of logical two-qubit interactions. Because the interplay between the physical and QEC layers is so profound, co-design methodologies are inherently technology-specific, tailored to exploit the unique properties and constraints of each qubit technology.

Several examples of this co-design philosophy have emerged across different qubit technologies. IBM's heavy-hexagon lattice was engineered specifically for the heavy-hexagon code~\cite{chamberland2020topological}, adopting a deliberately sparse connectivity that mitigates frequency collisions and crosstalk while remaining well-matched to the code's stabilizer structure. For superconducting systems, Duckering et al.~\cite{duckering2020virtualized} proposed a "2.5D" architecture that uses high-coherence cavity memories to virtualize logical qubits, enabling transversal CNOTs between logical qubits sharing the same physical address. In neutral atom platforms, Viszlai et al.~\cite{viszlai2025interleaved} designed a group-interleaved surface-code layout that exploits the platform's reconfigurability to enable fast transversal CNOTs within groups of logical qubits, reducing the spacetime overhead of logical operations compared to lattice surgery. For trapped-ion QCCD architectures, Jones and Murali~\cite{jones2026architecting} performed a QEC-aware co-design analysis in which hardware parameters such as trap size and communication topology are tuned specifically for the surface code, using a topology-aware compiler to quantify the performance implications of each architectural choice. In the shuttling domain, including spin and trapped-ion qubits, Cai, Siegel, and Benjamin~\cite{cai2023looped} introduced a looped-pipeline architecture in which planes of qubits are continuously shuttled in a loop, emulating 3D lattice connectivity on a strictly 2D device and enabling transversal gate execution among logical qubits.

Finally, and most directly relevant to the present work, Escofet et al.~\cite{escofet2026synthesizing} introduced Quantum Reverse Mapping for spin-qubit shuttling architectures: rather than mapping a fixed QEC code onto a predetermined hardware topology, this methodology derives the optimal hardware layout from the code's syndrome-extraction requirements, yielding a one-dimensional shuttling-bus architecture for the QEC code that minimizes both shuttling distance and cycle latency.

In this work, we extend the Quantum Reverse Mapping methodology beyond single-logical-qubit storage and syndrome extraction. We propose a full multi-qubit spin-qubit architecture that leverages coherent spin shuttling to support transversal two-qubit logical gates across pairs of logical qubits, while preserving the per-element optimality of the syndrome extraction design~\cite{escofet2026synthesizing}. We explore a range of architectural tradeoffs, including ancilla sharing across logical qubits, symmetric versus asymmetric shuttling policies, and 1D versus 2D bus topologies, and assess their impact on logical error rates for both memory and transversal CNOT operations. We further demonstrate the practical utility of the architecture by applying it to magic state distillation, showing that Quantum Reverse Mapping applied at the logical level yields optimized distillation circuits with minimal transveral CNOT (tCNOT) overhead. Our main contributions are as follows:

\begin{itemize}
    \item A multi-qubit shuttling bus architecture that enables all-to-all logical connectivity between logical qubits via transversal two-qubit gates, building upon the single-logical-qubit design of~\cite{escofet2026synthesizing} and extending it to an arbitrary number of concatenated logical elements.

    \item A comprehensive noise characterization of the proposed architecture under a realistic spin-qubit noise model, benchmarked for both memory and transversal CNOT experiments across a wide range of noise parameters and code distances.

    \item An ancilla-sharing scheme that encodes multiple logical qubits within a single logical element, reducing inter-qubit shuttling distances and improving tCNOT performance for long-range interactions, with a quantitative analysis of the resulting memory-computation tradeoff.

    \item An extension of the architecture to a 2D grid of shuttling tracks via lightweight junction elements, reducing the inter-qubit distance and yielding consistent improvements in tCNOT logical error rates.

    \item We assess how the transversal gate capabilities of the proposed architecture can be leveraged to implement the \textit{15-to-1} magic state distillation circuit. By applying the Quantum Reverse Mapping methodology at the logical level, we obtain an optimized logical qubit placement that minimizes tCNOT cost and naturally positions the distilled $T$ state for efficient injection into a larger processing unit.

\end{itemize}

The remainder of the paper is structured as follows. Section~\ref{sec:architecture} reviews the shuttling bus architecture and the Quantum Reverse Mapping methodology, and introduces transversal gates. Section~\ref{sec:multi_qubit_arch} describes the multi-qubit architecture proposed and its noise model, and benchmarks both the memory and transversal CNOT performance. Sections~\ref{sec:anc_qubit_sharing} and~\ref{sec:junctions} explore architectural tradeoffs (ancilla sharing and 2D connectivity) to enhance the performance of long-range transversal interactions. Section~\ref{sec:distillation} applies these architectural choices to the magic state distillation subroutine. Finally, Section~\ref{sec:discussion} discusses the implications of QEC code selection and the broader co-design landscape, and Section~\ref{sec:conclusions} presents the conclusions obtained in this work.

\section{Shuttling Bus Architecture}
\label{sec:architecture}

As the number of physical qubits on a spin-qubit chip grows, so it does the number of classical control lines needed to address them individually (DC, pulsed, or microwave) yet the chip's available wiring cross-section is bounded by its perimeter rather than its area. This mismatch produces a pin-count bottleneck that fundamentally limits scalability~\cite{vandersypen2017interfacing}. Crossbar~\cite{li2018crossbar} and bilinear~\cite{mohiyaddin2021large} layouts ease this fan-out problem without eliminating it.

One way around this constraint is to stop wiring every qubit individually and instead move qubits to a shared set of interaction sites as needed, using a 1-D "conveyor" track that shuttles qubits into position only when a gate is to be applied. Two proposals exemplify this idea. Patomäki \textit{et al.} designed a pipeline architecture that advances qubits column-by-column through fixed single- and two-qubit gate sites, driven by global timing signals~\cite{patomaki2024pipeline}. Künne \textit{et al.} designed SpinBus, a modular array of parallel shuttling lanes that achieves 2D logical connectivity using just four phase-shifted sinusoidal control voltages per lane~\cite{kunne2024spinbus}. Both build on experimental progress in coherent electron-spin shuttling~\cite{seidler2022conveyor, zwerver2023shuttling}.

We call this general class of designs the \emph{shuttling bus architecture}. It comprises two kinds of sites, linked by a common shuttling bus: \emph{operation zones}, dedicated locations where single- and two-qubit gates and readout are carried out, and \emph{storage zones}, where qubits sit idle as stationary quantum dots between operations.

Micromagnets placed above the gates of the \textit{operation zones} are required to drive single- and two-qubit operations (e.g., via electric-dipole spin resonance), but their stray magnetic field gradient also couples electric and charge noise into the qubit's spin degree of freedom, locally degrading its dephasing time whenever the qubit resides in their vicinity~\cite{philips2022universal, unseld2025baseband}. Conversely, \textit{storage zones} are kept free of micromagnets, allowing spins parked there to idle with comparatively long dephasing times. Qubits are moved between zones via the shuttling bus that connects all sites; since the bus implements a single global confinement potential, any qubits sharing a track segment must move in the same direction (right or left) and at the same speed. Quantum gates can only be applied within operation zones: every time a gate must be executed, the corresponding qubit (or, for two-qubit gates, both qubits) must first be shuttled from storage into a shared operation zone.

This general design space has been explored from several angles. A shuttle-based looped-pipeline scheme is used in~\cite{cai2023looped} to emulate 3D connectivity for the surface code on hardware that is strictly 2D. The SpinBus platform itself has been evaluated for surface-code implementations in~\cite{yenilen2025performance, chadwick2025manufacturable}. Separately, several works have developed compilation strategies specifically for shuttling-based hardware~\cite{escofet2025compilation, crawford2023compilation, paraskevopoulos2024besnake}.

Within a full-stack quantum computer, errors propagate upward from lower to higher layers of the stack, so it is essential to first guarantee reliable, error-corrected storage of a logical qubit via efficient syndrome extraction, before tackling computing primitives such as transversal gates. In this work, we take the design previously optimized for syndrome extraction in~\cite{escofet2026synthesizing} as our starting point and extend it to assess the computational-level implications of operating such an architecture.

\subsection{Storing: Quantum Reverse Mapping}
In~\cite{escofet2026synthesizing}, we introduced \emph{Quantum Reverse Mapping} as a methodology that, rather than mapping a fixed quantum error-correcting code onto a predetermined hardware connectivity, designs the hardware layout directly from the requirements of the code itself. Concretely, we used this methodology to synthesize a one-dimensional shuttling bus architecture optimized to host a single logical qubit encoded in the rotated surface code \cite{bravyi1998quantum,dennis2002topological,fowler2009high,fowler2012surface,litinski2019game}, with the explicit goal of making syndrome extraction as fast and as shuttling-efficient as possible.

The main idea is to encode the data-ancilla interaction pattern of one syndrome-extraction round as a \emph{chain-preserving directed acyclic graph} (DAG) \cite{thulasiraman2011graphs}, where each data qubit connects, in order, to the ancilla qubits it must interact with. Since this DAG is acyclic for the rotated surface code, any topological ordering gives a valid 1-D placement in which all data qubits can be shuttled in parallel without violating an interaction dependency. We optimized such placements with a Mixed-Integer Linear Program (MILP) \cite{bertsimas1997introduction} that jointly minimizes total shuttling distance and round cycle time. As the MILP becomes intractable for larger code distances ($O(d^4)$ variables), we proposed the \emph{Zig-Zag} heuristic, which mirrors the local structure of MILP-optimal solutions, scales linearly with the number of data qubits ($O(d^2)$), and matches the optimal placement exactly for the non-prohibitive code distances.

Compared to a naive topological placement, both the MILP-optimal and Zig-Zag layouts reduce shuttling distance and round duration from quadratic to linear scaling in $d$, and require only a constant number of extra idling quantum dots (two), rather than one growing linearly with $d$. Noise simulations combining gate depolarization, quantum-dot dephasing, and shuttling-induced dephasing confirmed logical error rates as low as $2\times10^{-10}$ per round at $d=21$, validating the architecture as a viable substrate for a single fault-tolerant logical qubit.

\begin{figure}
    \centering
    \includegraphics[width=1\linewidth]{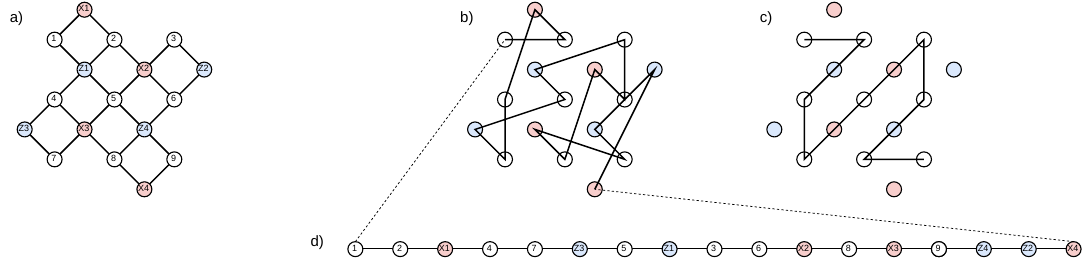}
    \caption{Quantum Reverse Mapping summary. \textbf{a)} Distance-3 Surface Code. \textbf{b)} Optimal relative placement obtained from the MILP model, expanding such order in a 1D results in the relative placement of each qubit in a single bus, depicted in \textbf{d)}. \textbf{c)} Depicts the same ordering as \textbf{b)} but omitting the ancilla qubits, giving name to the \textit{Zig-Zag} placement heuristic.}
    \label{fig:QRM_summary}
\end{figure}

Figure \ref{fig:QRM_summary} summarizes the output of the \textit{Quantum Reverse Mapping} synthesis method. Starting with a surface code patch, the qubits are placed in a 1D ordering that minimizes the shuttling time and distance needed for obtaining the QEC syndrome. 

\subsection{Computing: Transversal Gates}

Reliably storing a logical qubit is only half of the picture: the ultimate goal of having reliable, error-corrected qubits is to perform computation with them. This work extends the storage design proposed in~\cite{escofet2026synthesizing} by exploring how logical gates can be applied between error-corrected qubits.

A natural candidate, particularly popular in superconducting architectures given their nearest-neighbor connectivity, is Lattice Surgery (LS)~\cite{horsman2012surface}. In LS, logical two-qubit operations are implemented via a sequence of joint stabilizer measurements along a merge-split path connecting two logical patches, requiring $O(d)$ rounds of syndrome extraction (SE). While well-suited to 2D grid hardware, LS carries significant overheads: the $O(d)$ SE rounds introduce substantial latency, ancilla qubits must be reserved to route the merge-split path between patches, and logical gates sharing the same routing path must be executed sequentially.

A much more promising alternative is to apply logical two-qubit gates \emph{transversally}, where data qubits from two logical patches interact pairwise (i.e., data qubit $i$ from logical qubit $A$ interacts with data qubit $i$ from logical qubit $B$) in a single step, incurring no additional space or time overhead. The fundamental challenge of transversal gates is that they require physical interactions between data qubits that may reside in distant, spatially separated patches, demanding long-range connectivity. This is precisely why LS is the dominant paradigm in superconducting qubit architectures, where connectivity is inherently local, while transversal two-qubit gates remain out of reach. However, several architectures specifically engineered to support transversal logical gates have been proposed across different qubit modalities, including superconducting qubits~\cite{duckering2020virtualized}, neutral atoms~\cite{viszlai2025interleaved}, trapped ions~\cite{gutierrez2019transversality}, and spin qubits~\cite{cai2023looped, chadwick2025manufacturable}.

In this work, we propose an architecture that leverages spin-qubit shuttling to enable transversal logical two-qubit gates, while remaining co-designed with, and building directly upon, the layout previously optimized for robust logical qubit correction~\cite{escofet2026synthesizing}.

\section{A Multi-Qubit Architecture}
\label{sec:multi_qubit_arch}

\begin{figure}
    \centering
    \includegraphics[width=0.65\linewidth]{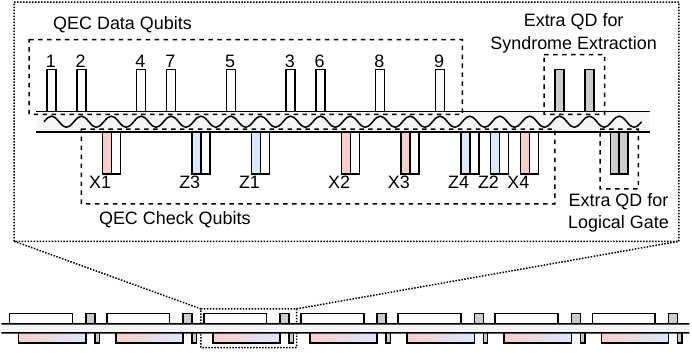}
    \caption{(\textbf{Top}) A single logical element for a distance-$d$ surface code: a 1D shuttling bus hosting $d^2$ storage zones (individual sites, upper side) and $d^2$ operation zones (paired sites, lower side). Two additional quantum dots are appended to the bus to ensure all spins can be reliably stored throughout a complete syndrome extraction cycle. (\textbf{Bottom}) A multi-qubit architecture formed by concatenating multiple logical elements along a single shuttling bus.}
    \label{fig:logical_element}
\end{figure}

The design output of the Quantum Reverse Mapping methodology applied to the surface code is a 1D array of data qubits interleaved with ancilla qubits~\cite{escofet2026synthesizing}. Two additional quantum dots must be appended to this array to reliably store all spins throughout a complete SE cycle. The resulting structure is depicted in the top row of Figure~\ref{fig:logical_element}, where \textit{storage zones} appear as individual sites on the upper side of the shuttling bus, and \textit{operation zones} appear as paired sites on the lower side, each accommodating a data and ancilla qubit respectively.

The pairwise physical interactions between data qubits of two logical patches, needed for a transversal logical interaction, are carried out in the operation zones, since micromagnets are required there to apply the desired qubit rotations. Because transversal gates are performed in between QEC cycles, the ancilla qubits occupying the operation zones hold trivial states and pose no issue. However, the SE-focused design of~\cite{escofet2026synthesizing} contains $d^2$ storage zones but only $d^2-1$ operation zones for a distance-$d$ surface code, due to the number of ancilla qubits in a surface-code patch. For all $d^2$ pairs of data qubits to interact simultaneously during a transversal gate, one additional operation zone must be introduced, bringing the number of operation zones to match the number of storage zones. This yields the design shown in the top row of Figure~\ref{fig:logical_element}, which we refer to as a \textit{logical element}.

Each logical element encapsulates all the infrastructure needed to store and protect a single logical qubit. To support transversal two-qubit gates, we must move beyond the single-logical-qubit design of~\cite{escofet2026synthesizing} and consider an architecture that hosts multiple logical elements. Constrained by the 1D nature of the shuttling bus, we do so by simply concatenating logical elements end to end along a single bus, as shown in the bottom row of Figure~\ref{fig:logical_element}.

In this multi-element design, each QEC cycle operates entirely within the domain of a single logical element, so data qubits need only shuttle the minimal distances established in~\cite{escofet2026synthesizing}. At the same time, every logical qubit is within reach of every other, yielding an all-to-all logical connectivity. Whenever two logical qubits must interact, their respective data qubits are simultaneously shuttled pairwise into operation zones, where a two-qubit gate is applied, after which each qubit returns to its assigned storage zone.

\begin{figure}
    \centering
    \includegraphics[width=0.5\linewidth]{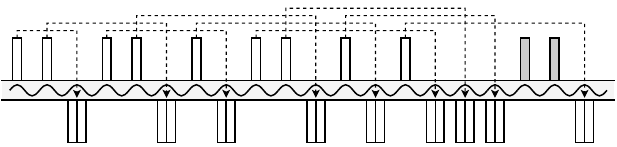}
    \caption{Shuttling assignments for a transversal two-qubit gate between two logical elements. The $i$-th leftmost data qubit of each logical element is assigned to the $i$-th leftmost operation zone, minimizing the total shuttling distance from storage to operation zones for the local logical element. The second logical qubit involved (not depicted in the figure) follows the same pairing rule, though it will come from distant storage zones.}
    \label{fig:shuttling_storage_operation}
\end{figure}

Since each logical element now contains exactly $d^2$ storage zones and $d^2$ operation zones, we assign one operation zone to each of the $d^2$ data qubits, with the goal of minimizing the total shuttling distance from storage to operation zones. To this end, the qubit in the $i$-th leftmost storage zone is assigned to the $i$-th leftmost operation zone, as illustrated in Figure~\ref{fig:shuttling_storage_operation}. For the second logical qubit involved in the gate, the same pairing rule applies, though it necessarily incurs a longer shuttle to reach its designated operation zones. The effect of different shuttling policies for transversal interactions is examined in detail in later sections.

These transversal-gate shuttling schedules should not be confused with the SE shuttling schedules required during a QEC cycle, which are dictated by the code structure and involve multiple sequential shuttling steps interleaved with stabilizer measurements. A complete characterization of the space-time overhead of SE shuttling is provided in~\cite{escofet2026synthesizing}.

\subsection{Noise Model}

We adopt a realistic noise model similar to that used in~\cite{escofet2026synthesizing}. Since we are assessing the viability of a novel spin-qubit architecture, the noise model must reflect the distinct error mechanisms present at each stage of the computation, rather than relying on a single high-level error mechanism.

To this end, we distinguish four sources of noise:
\begin{itemize}
    \item \textbf{Gate noise:} Whenever a physical gate is applied, the involved qubit(s) undergo depolarization at a rate proportional to the gate error rate. For two-qubit gates, both qubits are subject to joint depolarization.
    \item \textbf{Readout noise:} Measurement errors are modeled as a Pauli-$X$ channel applied immediately before each measurement, leading to an erroneous readout outcome with probability proportional to the channel strength.
    \item \textbf{Idling noise:} Qubits that are not actively participating in a gate or shuttle undergo relaxation and dephasing, modeled as Pauli-$X$ and Pauli-$Z$ noise channels governed by $T_1$ and $T_\phi$, respectively. Since the presence of micromagnets in the operation zones significantly reduces the local dephasing time~\cite{philips2022universal, unseld2025baseband}, we define two distinct dephasing times: ${T_\phi}^{\texttt{Store}}$ for qubits idling in storage zones, and ${T_\phi}^{\texttt{Op}}$ for qubits idling in operation zones, with ${T_\phi}^{\texttt{Store}} \gg {T_\phi}^{\texttt{Op}}$.
    \item \textbf{Shuttling noise:} Noise incurred during a shuttle is modeled as idling noise with an effective dephasing time that, following experimental observations~\cite{struck2024spin}, scales with the shuttled distance as ${T_\phi}^{\texttt{Sh}} = {T_\phi}^{\texttt{Bus}} \sqrt{\frac{d_{\texttt{Sh}}+l_c}{l_c}}$ where $l_c$ is the correlation length of the quantum dot array~\cite{struck2024spin}.
\end{itemize}

This noise model enables a realistic evaluation of the proposed architecture. Table~\ref{tab:parameters} lists the noise values and parameter ranges used throughout the remainder of the paper, derived from rounded estimates of state-of-the-art spin-qubit processors~\cite{Stano}; the shuttling velocity was estimated as a rounded value of the highest velocity in~\cite{desmet2025, li2026}, and $l_c$ was taken from~\cite{struck2024spin}.

\begin{table}
    \caption{Physical noise values and parameter ranges}
    \centering
    \begin{tabular}{c|c}
        \quad\quad\quad\quad \textbf{Parameter} \quad\quad\quad\quad & \quad\quad \textbf{Value} \quad\quad \\
        \hline
        \hline Single-qubit gate error & $1 \cdot 10^{-4}$\\
        \hline Two-qubit gate error & $5 \cdot 10^{-4}$\\
        \hline Measurement error & $1 \cdot 10^{-3}$\\
        \hline
        \hline Single-qubit gate time & $100$ ns\\
        \hline Two-qubit gate time & $50$ ns\\
        \hline Measurement time & $500$ ns\\
        \hline
        \hline $T_1$ & $500$ ms\\
        \hline $T_{\phi}^\texttt{Store}$ & [$100$ -- $800$] µs\\
        \hline $T_{\phi}^\texttt{Op}$ & $30$ µs\\
        \hline $T_{\phi}^\texttt{Bus}$ & [$50$ -- $1600$] µs\\
        \hline
        \hline $l_c$ & $13$ nm\\
        \hline Distance between QD & $100$ nm\\
        \hline Shuttling Velocity & $100$ m/s\\
    \end{tabular}
    \label{tab:parameters}
\end{table}

The noise model underlying spin qubits is strongly biased toward $Z$ errors, which are far more prevalent than $X$ errors. To illustrate this, we first perform a memory experiment (i.e., storing a single logical qubit over several QEC rounds) under a fixed choice of noise parameters\footnote{For this initial experiment we set $T_{\phi}^\texttt{Store} = T_{\phi}^\texttt{Bus} = 100 \mu\text{s}$ and $T_{\phi}^\texttt{Op} = 30 \mu\text{s}$, deferring a full parameter sweep to later sections.}.

\begin{figure}
    \centering
    \includegraphics[width=0.5714\linewidth]{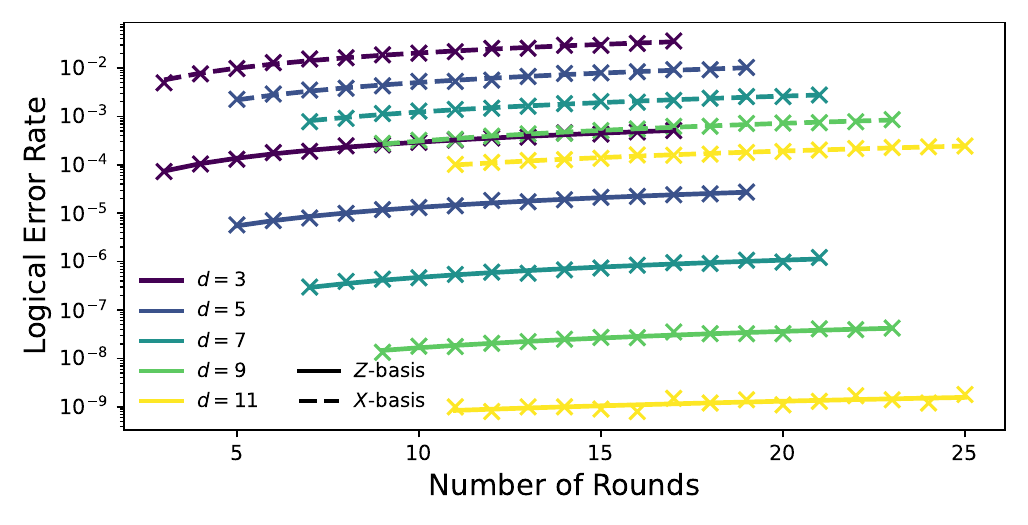}
    \caption{LER for a logical qubit memory experiment, shown for code distances $d=3$ to $d=11$. Results are reported for both the $Z$ basis and the $X$ basis. The strongly biased nature of the spin-qubit noise model, where Pauli-$Z$ errors dominate, leads to a substantially lower LER in the $Z$ basis across all code distances. Noise parameters are fixed to $T_{\phi}^\texttt{Store} = T_{\phi}^\texttt{Bus} = 100 \mu s$ and $T_{\phi}^\texttt{Op} = 30 \mu s$.}
    \label{fig:both_bases_memory}
\end{figure}

Figure~\ref{fig:both_bases_memory} shows the logical error rate (LER) as a function of the number of SE rounds for code distances $d=3$ to $d=11$, measured in both the $Z$ and $X$ bases (solid and dashed lines respectively). In the $Z$ basis, the logical state is prepared as $|0\rangle_L$, which commutes with Pauli-$Z$ noise, and consequently achieves a substantially lower LER. In the $X$ basis, the state is prepared as $|+\rangle_L$, which commutes with $X$ but not with $Z$ noise; since $Z$ errors dominate in our model, the LER in the $X$ basis is orders of magnitude higher than in the $Z$ basis.

A complete characterization would benchmark both bases and report an average LER. However, given the large performance gap between the two in our setting, we instead adopt the $X$ basis as a worst-case assessment of the architecture and report only its LER throughout the remainder of the paper. Accordingly, the figures that follow show not an estimated average, but the most pessimistic LER achievable, for both the memory and transversal CNOT experiments.

\subsection{Memory Capabilities}

Given the rapid and continued progress in spin coherence times and shuttling fidelity, rather than fixing the dephasing times $T_{\phi}^\texttt{Store}$ and $T_{\phi}^\texttt{Bus}$ to a single point, we explore a range of values representative of current and near-future devices. As summarized in Table~\ref{tab:parameters}, $T_{\phi}^\texttt{Store}$ is swept from a conservative $100 \mu s$ to an optimistic yet achievable $800 \mu s$, and the shuttling-bus 
dephasing time is coupled to it via $T_{\phi}^\texttt{Bus} = \alpha\, T_{\phi}^\texttt{Store}$, 
with $\alpha \in \{\frac{1}{2}, 1, 2\}$.

\begin{figure}
    \centering
    \includegraphics[width=\linewidth]{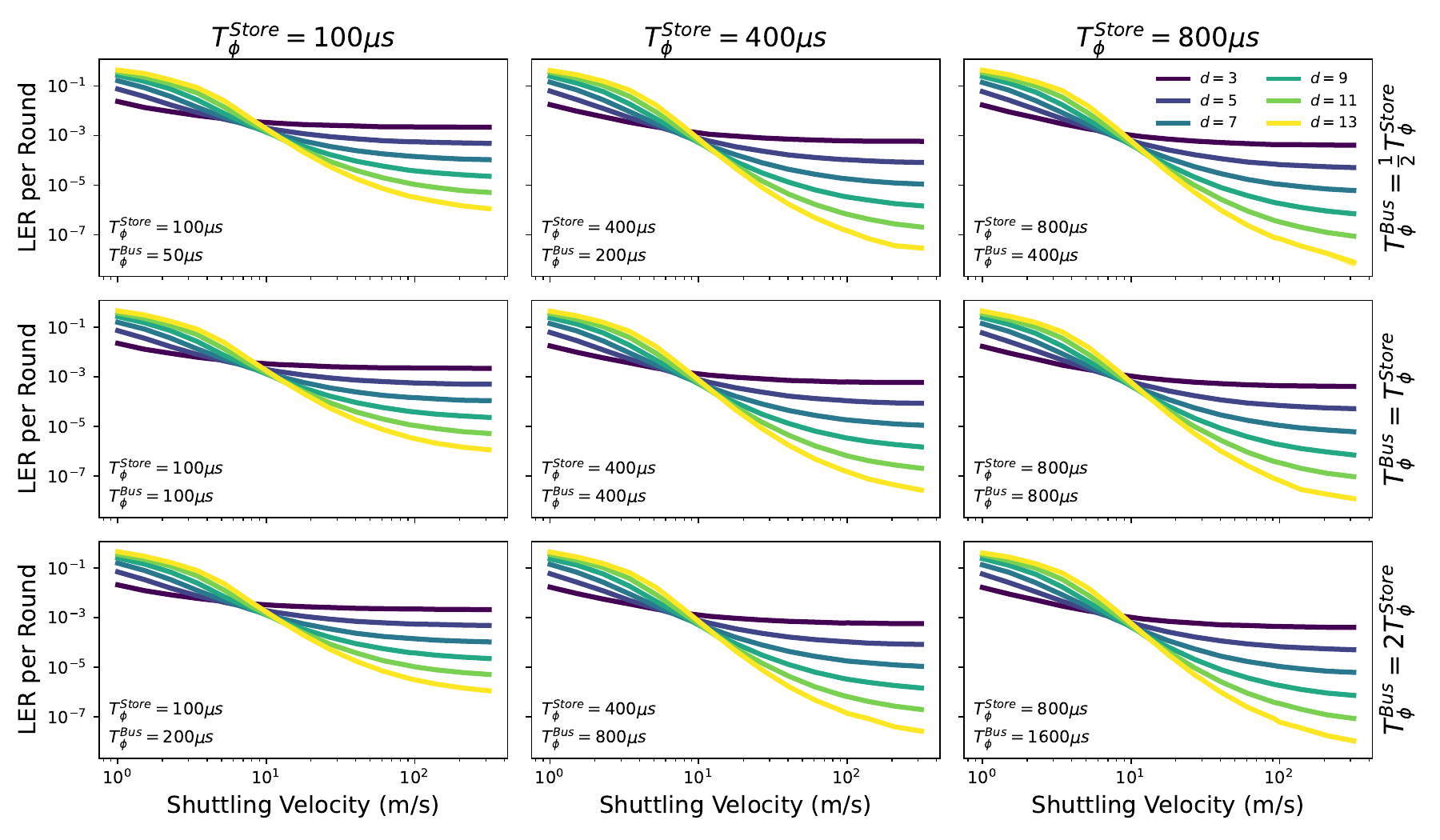}
    \caption{LER per SE round, shown for code distances $d = 3$ to $d = 13$ as a function of shuttling velocity ($1$ to $320$ m/s). Each panel corresponds to a distinct combination of storage-zone dephasing time $T_{\phi}^\texttt{Store} \in \{100, 400, 800\} \mu s$ and shuttling-bus dephasing time $T_{\phi}^\texttt{Bus} = \alpha\, T_{\phi}^\texttt{Store}$, with $\alpha \in \{\frac{1}{2}, 1, 2\}$. The threshold velocity, below which increasing the code distance no longer reduces the LER, is consistent across all noise configurations, ranging from $9$ to $9.5$ m/s.}
    \label{fig:memory_velocity}
\end{figure}

Figure~\ref{fig:memory_velocity} shows, for each combination of $T_{\phi}^\texttt{Store}$ and $T_{\phi}^\texttt{Bus}$, the LER per error-correction round as a function of shuttling velocity, swept from $1$ to $320$ m/s. The per-round LER $\mathcal{E}_d$ is extracted by measuring the logical failure probability $P_L$ over an increasing number of SE rounds $n \in [d,\, d+20]$ and fitting the data to~\cite{lacroix2025scaling}:
\begin{equation}
    P_L = \mathcal{E}_0(1-\mathcal{E}_d)^n + \frac{1}{2}
\end{equation}
with fitting parameters $\mathcal{E}_0$ and $\mathcal{E}_d$.

This fitting separates the time-dependent logical decay from static overheads: $\mathcal{E}_0$ captures errors that do not accumulate with the number of SE cycles, while $\mathcal{E}_d$ represents the probability of a logical error occurring within a single SE cycle, the value reported in 
Figure~\ref{fig:memory_velocity}.

As expected, shuttling velocity has a strong impact on the LER across all tested noise configurations and code distances: higher velocities reduce qubit idle time per SE round, directly lowering the accumulated dephasing error. Regarding the dephasing time exploration, increasing $T_{\phi}^\texttt{Store}$ from $100$ to $400 \mu s$ yields a substantial improvement in LER, whereas a further increase to $800 \mu s$ brings only marginal gains. This saturation indicates that $T_{\phi}^\texttt{Store}$ is no longer the limiting noise source at higher values, and that performance is instead constrained by other channels, such as the dephasing time in operation zones or gate depolarization. We also observe that $T_{\phi}^\texttt{Bus}$ has a negligible impact on the LER across all values considered.

An important practical question is at which shuttling velocity the architecture is within QEC threshold that is, the regime where increasing the code distance consistently decreases the LER. This threshold velocity is identified by the crossing point between LER curves at different code distances, which we find to be remarkably stable across noise configurations, ranging from $9$ to $9.5$ m/s.

\begin{figure}
    \centering
    \includegraphics[width=\linewidth]{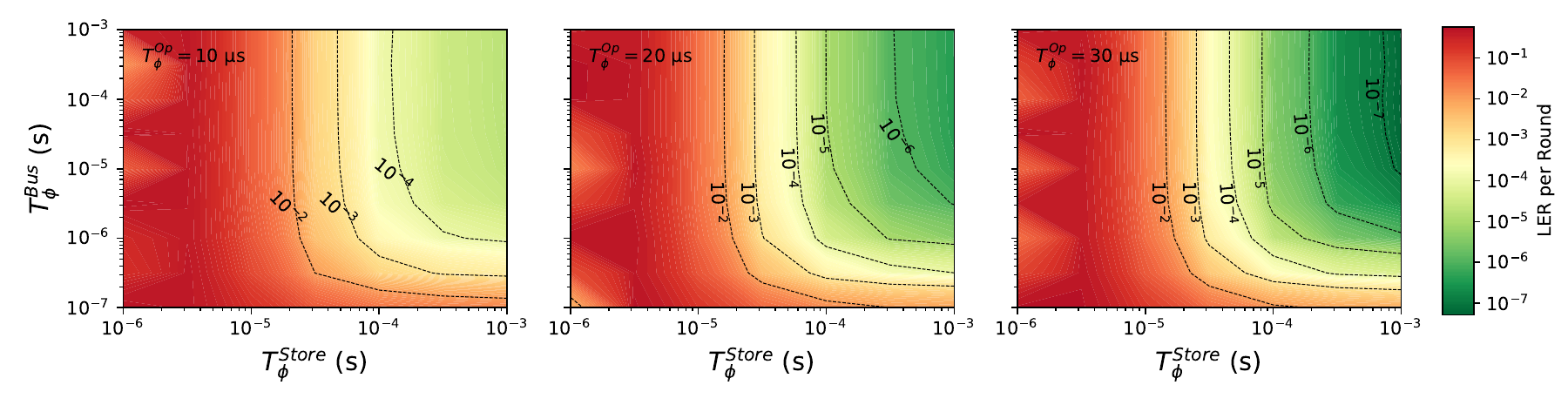}
    \caption{LER per SE round at fixed code distance $d = 13$ and shuttling velocity $v_\texttt{Sh} = 100$ m/s, as a function of storage-zone dephasing time $T_{\phi}^\texttt{Store}$ and shuttling-bus dephasing time $T_{\phi}^\texttt{Bus}$. Results are shown for three discrete values of the operation-zone dephasing time $T_{\phi}^\texttt{Op} \in \{10, 20, 30\} \mu s$.}
    \label{fig:memory_sweep}
\end{figure}

To further assess the contributions of the individual dephasing sources more clearly, we fix the code distance to $d = 13$ and the shuttling velocity to $v_\texttt{Sh} = 100$ m/s and perform a finer sweep over $T_{\phi}^\texttt{Store}$ and $T_{\phi}^\texttt{Bus}$. Figure~\ref{fig:memory_sweep} reports the resulting LER for three discrete values of the operation-zone dephasing time: $T_{\phi}^\texttt{Op} \in \{10, 20, 30\} \mu s$.


Even a modest increase in $T_{\phi}^\texttt{Op}$ produces a pronounced improvement in LER across all tested configurations, confirming that the operation-zone dephasing time is the dominant limiting factor in error-correction performance. The sweep also reveals that $T_{\phi}^\texttt{Bus}$ saturates early, while $T_{\phi}^\texttt{Store}$ exerts a broader influence on the LER over the full range explored.

\subsection{Transversal CNOT}

We now benchmark the defining feature of the proposed architecture: transversal logical interactions between logical qubits. To this end, we employ a Logical Randomized Benchmarking (LRB) methodology~\cite{combes2017logical, lacroix2025scaling}, which applies a sequence of $n$ tCNOT gates, each interleaved with a round of error correction, and extracts the per-gate LER from the decay of logical fidelity as a function of sequence length $n$. The measured LER is fitted to a two-qubit analog of the expression used 
for the memory experiment:
\begin{equation}
    P_L = \mathcal{E}_0(1-\frac{4}{3}\mathcal{E}_{cx})^n + \frac{3}{4}
\end{equation}
where $\mathcal{E}_{cx}$ is the per-tCNOT LER reported throughout this section.

The dominant source of error in a tCNOT is expected to be the physical distance between the two involved logical qubits. As shown in Figure~\ref{fig:logical_element}, logical elements are concatenated along a single shuttling bus, so for two logical qubits to interact transversally, their respective data qubits must first be shuttled to a shared set of operation zones and then returned to their assigned storage positions.

Drawing on insights from prior work on shuttling-bus compilation~\cite{escofet2025compilation}, the policy that minimizes overall circuit latency (and therefore accumulated error) consists of shuttling all qubits simultaneously in the same direction, both before and after the gate. In the proposed architecture, this translates to the following schedule. Suppose logical qubit $A$, stored at logical element $i$, must interact with logical qubit $B$, stored at logical element $j > i$. Both logical qubits are simultaneously shuttled to the right and brought together in the operation zones of logical element $j$, where the pairwise data-qubit interactions take place; after the gate, both qubits are simultaneously shuttled to the left. We define the initial separation 
between the two logical qubits as the number of \textit{logical hops}, i.e., the number of 
logical elements that the leftmost qubit must traverse to reach the operation zones of the 
rightmost element.

\begin{figure}
    \centering
    \includegraphics[width=0.85\linewidth]{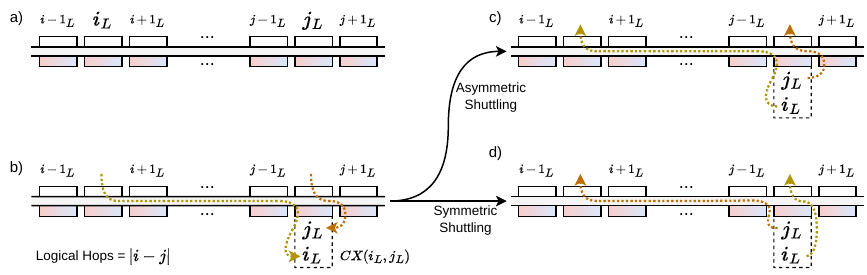}
    \caption{Shuttling schedules for a tCNOT between logical qubits stored at positions $i$ and $j$, with $i < j$. \textbf{a)} and \textbf{b)} depict the forward shuttling step, in which both logical qubits are simultaneously moved to the right and brought together in the operation zones of logical element $j$, where all $d^2$ pairs of data qubits interact pairwise. Qubit $i$ travels $|i-j|$ logical hops to reach the designated operation zones, while qubit $j$ covers only the short intra-element distance between its storage and operation zones. \textbf{c)} \textit{asymmetric shuttling}, in which each qubit is returned to its original storage position. \textbf{d)} \textit{symmetric shuttling}, in which the storage positions of the two logical qubits are swapped after the interaction, so that each qubit travels the same total number of logical hops across the full gate schedule.}
    \label{fig:sym_shuttling}
\end{figure}

Figures~\ref{fig:sym_shuttling}a and~\ref{fig:sym_shuttling}b depict this forward shuttling step for logical qubits stored at positions $i_L$ and $j_L$ with $i < j$: qubit $i_L$ is shuttled $|i - j|$ logical hops to reach the operation zones within logical element $j$, where both sets of data qubits interact pairwise.

Panels c) and d) of Figure~\ref{fig:sym_shuttling} illustrate two alternative strategies for returning the data qubits to storage zones after the interaction. In panel c), each logical qubit returns to its original position: qubit $i_L$ is shuttled back by as many logical hops as it traveled forward, while qubit $j_L$ only needs to cover the short distance between the operation and storage zones within its own logical element (see Figure~\ref{fig:shuttling_storage_operation}). This creates a pronounced asymmetry in the amount of shuttling (and therefore noise) accumulated by each qubit. We refer to this approach as \textit{asymmetric shuttling}.

Panel d) shows an alternative motivated by the compilation results of~\cite{escofet2025compilation}: by swapping the storage positions of logical qubits $i_L$ and $j_L$ after the interaction, both qubits travel the same total number of logical hops, one qubit before the gate and the other one after. We refer to this as \textit{symmetric shuttling} and expect it to yield a lower LER than the asymmetric alternative, since the noise burden is distributed more evenly between the two qubits, reducing the probability that one accumulates enough errors to exceed the correction capacity of the code.

\begin{figure}
    \centering
    \includegraphics[width=\linewidth]{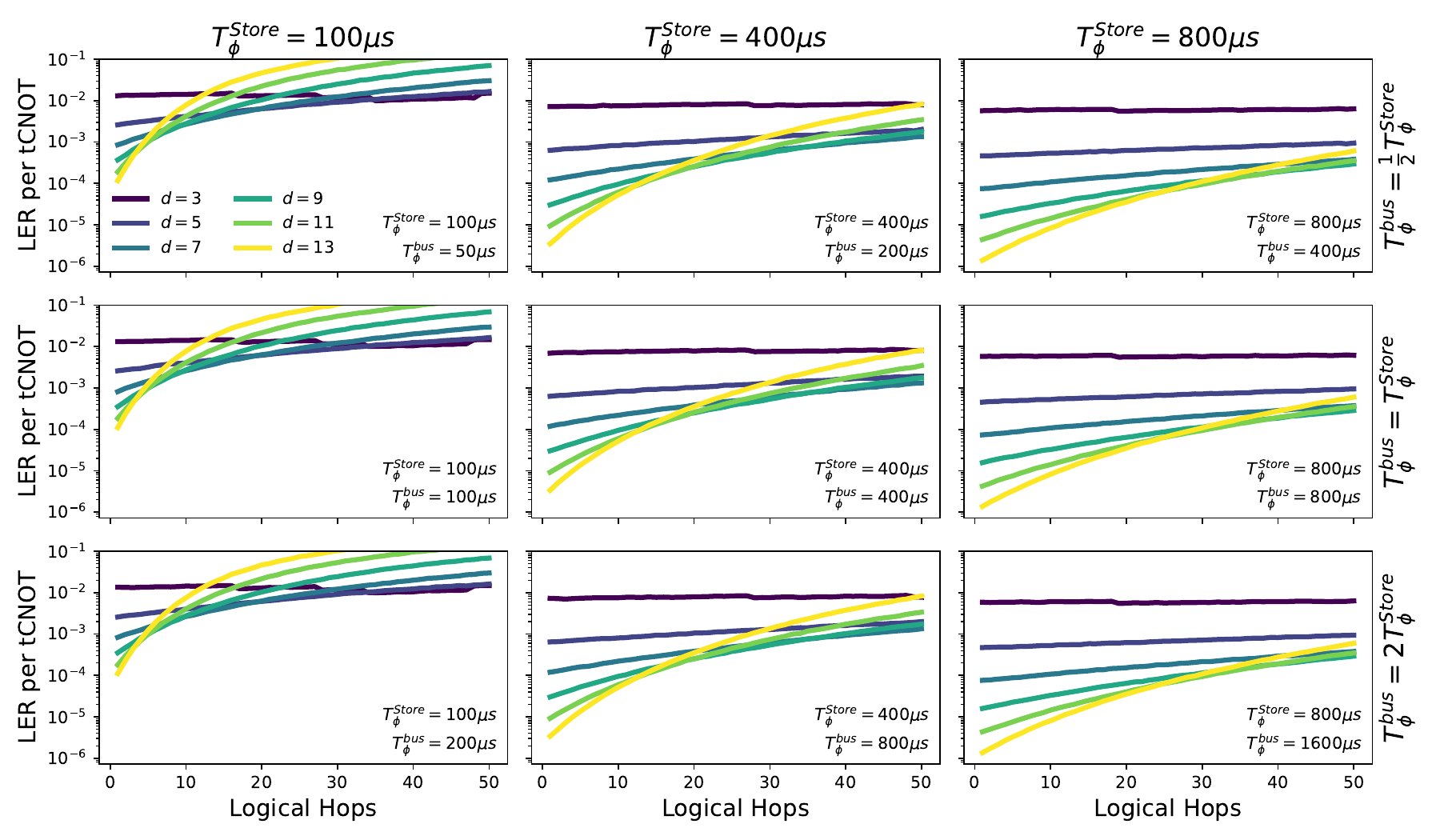}
    \caption{LER per tCNOT as a function of the number of logical hops separating the two interacting logical qubits, for code distances $d = 3$ to $d = 13$. Each panel corresponds to a distinct combination of storage-zone dephasing time $T_{\phi}^\texttt{Store} \in \{100, 400, 800\} \mu s$ and shuttling-bus dephasing time $T_{\phi}^\texttt{Bus} = \alpha\, T_{\phi}^\texttt{Store}$, with $\alpha \in \{\frac{1}{2}, 1, 2\}$.}
    \label{fig:transv_cnot}
\end{figure}

Figure~\ref{fig:transv_cnot} reports, for the same noise configurations as the memory experiment in Figure~\ref{fig:memory_velocity}, the LER of a single tCNOT (followed by one SE round) as a function of the number of logical hops, at fixed shuttling velocity $v_\texttt{Sh} = 100$ m/s and operation-zone dephasing time 
$T_{\phi}^\texttt{Op} = 30 \mu s$.

The results mirror the trends observed in the memory experiment. Increasing $T_{\phi}^\texttt{Store}$ from $100$ to $400 \mu s$ produces a consistent and significant reduction in LER across all logical hop counts. A further increase to $800 \mu s$ yields negligible improvement for short-distance interactions (1 to 10 logical hops), but its benefit grows with qubit distance, reaching up to an order of magnitude improvement at 50 logical hops for $d = 13$. As before, $T_{\phi}^\texttt{Bus}$ has a negligible impact across all noise configurations.

Notably, in Figure~\ref{fig:transv_cnot}, we see curves for different code distances cross at specific logical hop counts, revealing a clear tradeoff between error-correction capability and interaction distance. Increasing the code distance $d$ improves the error-correction capabilities, allowing the code to identify and correct a greater number of errors, but also increases the number of physical qubits per logical element and, consequently, the physical length of each logical element. As a result, shuttling through a fixed number of logical hops at larger code distances corresponds to a longer physical shuttle. Higher code distances therefore exhibit a steeper LER growth with logical hops, reflecting this code-distance tradeoff for long-range logical interactions.

\begin{figure}
    \centering
    \includegraphics[width=\linewidth]{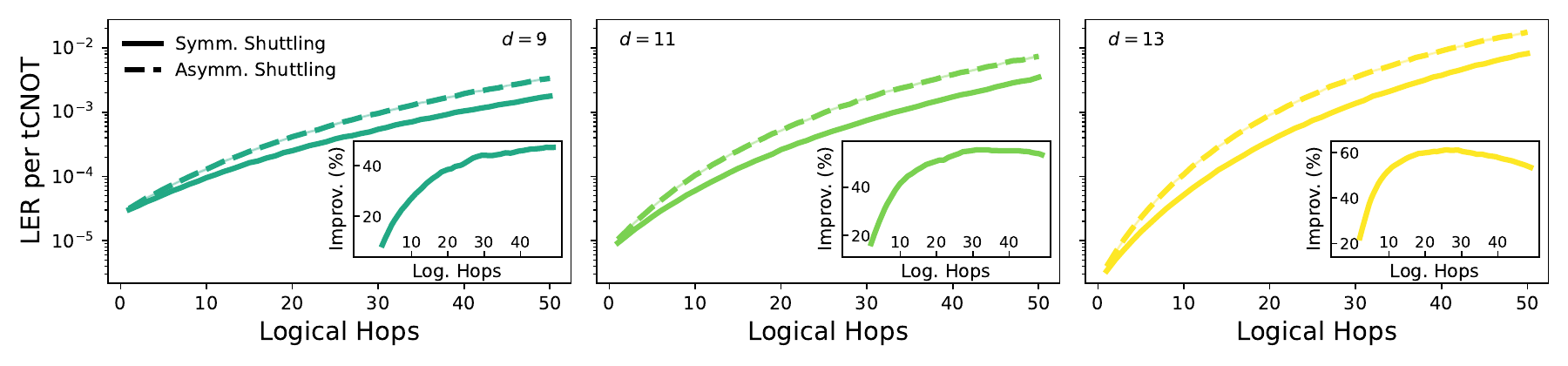}
    \caption{Comparison of asymmetric and symmetric shuttling strategies, shown for code distances $d = 9$, $11$, and $13$ as a function of the number of logical hops. Symmetric shuttling (solid lines) consistently achieves a lower LER than asymmetric shuttling (dashed lines) across all configurations, with improvements of up to $45\%$, $50\%$, and $60\%$ for $d = 9$, $11$, and $13$, respectively. Inset plots show the improvement percentage as a function of logical hops for each code distance.}
    \label{fig:transv_cnot_sym}
\end{figure}

We now verify the claim that symmetric shuttling outperforms the asymmetric alternative by directly comparing the two strategies under a single representative noise configuration\footnote{We fix $T_{\phi}^\texttt{Store} = T_{\phi}^\texttt{Bus} = 400 \mu s$ and $T_{\phi}^\texttt{Op} = 30 \mu s$; the same qualitative conclusions hold across the full range of noise parameters explored.}. Figure~\ref{fig:transv_cnot_sym} shows the LER per tCNOT as a function of logical hops for code distances $d = 9$, $11$, and $13$. Symmetric shuttling consistently outperforms asymmetric shuttling across all configurations, with improvement percentages reaching up to $45\%$, $50\%$, and $60\%$ for $d = 9$, $11$, and $13$, respectively, as shown in the inset of each panel.

For $d = 11$ and $d = 13$, the improvement percentage eventually converges and then decreases at large logical hop counts. This behavior arises because the LER of a two-qubit system is bounded above by $3/4$. The asymmetric strategy saturates at this bound earlier, causing its curve to plateau while the symmetric strategy (which degrades more slowly) continues to grow, thereby narrowing the relative gap between the two.

The results presented so far demonstrate competitive LERs for both the memory experiment and transversal CNOT interactions. However, the LER of a tCNOT is strongly dependent on the number of logical hops separating the two interacting qubits, a critical feature for large-scale architectures potentially hosting thousands of logical qubits. In the following sections, we explore architectural tradeoffs between memory fidelity and logical connectivity, seeking to reduce the effective cost of long-distance transversal interactions.

\section{Ancilla-Qubit Sharing}
\label{sec:anc_qubit_sharing}

Despite its many advantages, one of the principal drawbacks of the surface code is its unfavorable physical-to-logical qubit ratio: encoding a single logical qubit requires $d^2$ data qubits and $d^2 - 1$ ancilla qubits, the latter dedicated entirely to syndrome extraction. In hardware with fixed nearest-neighbor connectivity, such as superconducting qubit grids, this overhead is unavoidable, necessitating $\sim 2 d^2$ physical qubits per logical qubit~\cite{google2023suppressing, google2025quantum}.

The all-to-all logical connectivity inherent to the proposed shuttling-bus architecture relaxes this constraint. Because qubits can be brought together on demand, multiple logical qubits can share a single set of ancilla qubits for syndrome extraction, at the cost of some reduction in SE throughput. This opens the door to encoding several logical qubits within a single logical element, reducing the physical footprint per logical qubit and, crucially, the physical distance between logical qubits hosted within the same element.

Concretely, we increase the number of storage zones in a logical element while keeping the number of operation zones (and therefore ancilla qubits) fixed. The result is that multiple logical qubits share the same ancilla resources, and their syndrome extraction rounds must be performed sequentially rather than in parallel. Each logical qubit therefore incurs additional latency per QEC cycle, waiting for the other co-hosted qubits to be corrected before its own syndrome is extracted. The expected outcome is a tradeoff: memory performance per QEC round degrades due to the increased cycle time, but the reduced inter-qubit distance lowers the shuttling overhead of transversal CNOTs, improving their LER. The balance between these two effects will, in general, depend on the code distance.

\begin{figure}
    \centering
    \includegraphics[width=\linewidth]{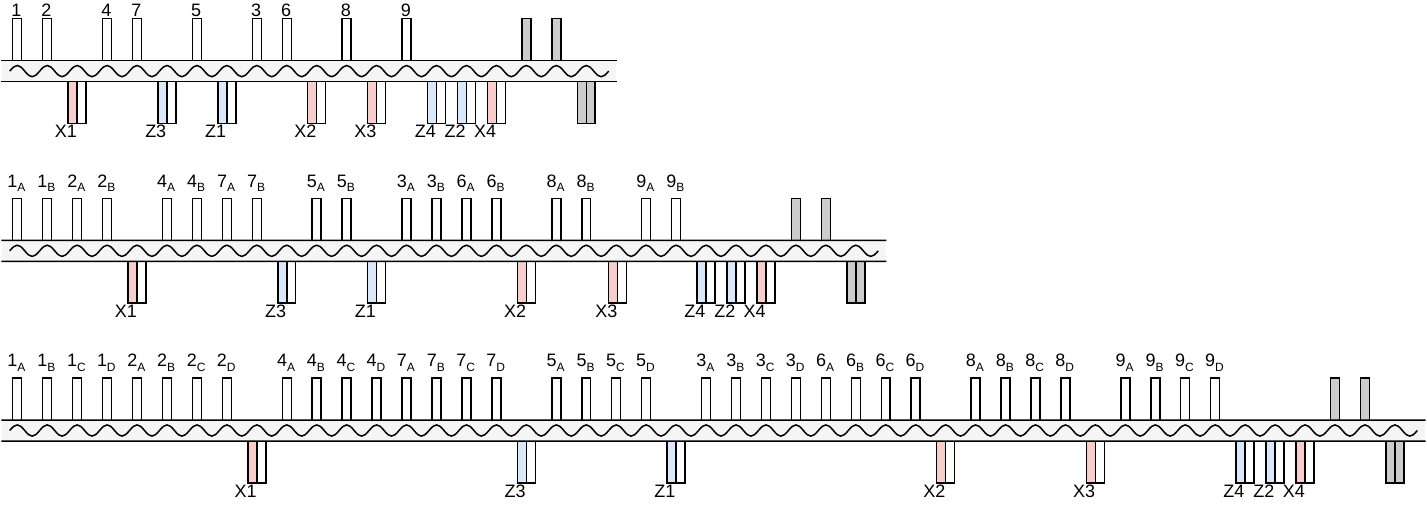}
    \caption{Logical element layouts for three encoding ratios $m$, shown for the distance-$3$ surface code. \textbf{(Top)} $m = 1$, baseline design, in which a single logical qubit is hosted per logical element, with $d^2$ storage zones and $d^2 - 1$ ancilla qubits dedicated to syndrome extraction. \textbf{(Middle)} $m = 2$, two logical qubits ($A$ and $B$) share a single set of $d^2 - 1$ ancilla qubits within one logical element. \textbf{(Bottom)} $m = 4$, four logical qubits ($A$, $B$, $C$, and $D$) share the same ancilla resources, each with 9 data qubits. The length overhead of a logical element with encoding ratio $m$ scales as $\sim \frac{m+1}{2}$.}
    \label{fig:ancilla_sharing}
\end{figure}

Figure~\ref{fig:ancilla_sharing} illustrates the logical element layout for three encoding ratios $m$ applied to the distance-$3$ surface code. At $m = 1$ (top row), the baseline design hosts a single logical qubit per logical element, as used throughout the preceding sections. At $m = 2$ and $m = 4$ (second and third rows), two and four logical qubits respectively share a single set of $d^2 - 1$ ancilla qubits within one logical element.

The space efficiency of higher encoding ratios is evident. Doubling the number of hosted logical qubits ($m = 2$) increases the logical element length by only a factor of $1.45\times$, encoding twice the logical information at a modest spatial overhead. Increasing to $m = 4$ yields four times the logical information at only $2.35\times$ the element length, demonstrating that ancilla sharing can substantially improve the logical information density of the shuttling bus.

With higher encoding ratios, we are able to encode more logical information without proportionally increasing the size of the logical element. By doubling the logical qubits within a logical element ($m=2$), we can store $2\times$ the information with a $\sim 1.5 \times$ length increase. And with $m=4$, comes a $4\times$ more logical information with a $\sim 2.5 \times$ logical element increase. More generally, for an arbitrary encoding ratio $m$, the length overhead of a logical element relative to the $m = 1$ baseline scales as $\sim \frac{m+1}{2}$, confirming the 
sub-linear growth in physical footprint with respect to the number of hosted logical qubits.

\subsection{Impact on Syndrome Extraction}

Sharing ancilla qubits across logical qubits introduces two compounding overheads on SE. First, data qubits must traverse longer shuttling distances to reach the operation zones, accumulating additional shuttling-induced dephasing. Second, the syndromes of logical qubits sharing the same ancilla set cannot be extracted simultaneously: each logical qubit must wait while the others are being corrected, incurring extended idling times in the storage zones and the associated decoherence. Together, these two effects cause the memory LER to increase relative to the $m = 1$ baseline, showcasing the fundamental tradeoff between memory performance and the connectivity gains offered by higher encoding ratios.

From the noise configurations explored in the preceding sections, we select two representative operating points: an already achievable baseline of $T_{\phi}^\texttt{Store} = 400 \mu s$ and a near-future target of $T_{\phi}^\texttt{Store} = 800 \mu s$. Given the negligible influence of $T_{\phi}^\texttt{Bus}$ observed in previous simulations, we fix $\alpha = 1$, setting $T_{\phi}^\texttt{Bus} = T_{\phi}^\texttt{Store}$ throughout. Narrowing the parameter space in this way allows us to present results more clearly and to draw conclusions that generalize readily to other noise regimes.

\begin{figure}
    \centering
    \includegraphics[width=\linewidth]{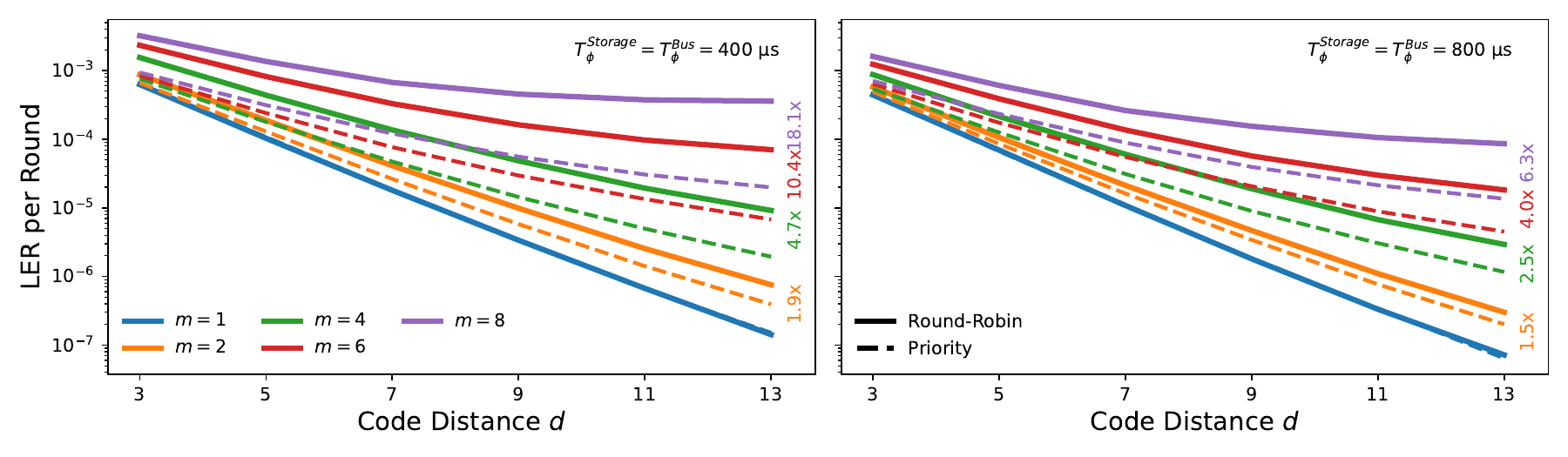}
    \caption{LER per SE round under encoding ratios $m \in \{1, 2, 4, 6, 8\}$, for code distances $d = 3$ to $d = 13$. Solid lines correspond to Round-Robin SE, in which syndrome extraction is performed sequentially for each logical qubit sharing the ancilla set. Dashed lines correspond to Priority SE, which corrects only one logical qubit per round. Annotations indicate the LER improvement of Priority over Round-Robin SE at $d = 13$ for each encoding ratio.}
    \label{fig:memory_ratio}
\end{figure}

Figure~\ref{fig:memory_ratio} shows the LER per SE round for code distances $d = 3$ to $d = 13$ under encoding ratios from $m = 1$ to $m = 8$, and under two distinct SE scheduling strategies. Solid lines correspond to \textit{Round-Robin SE}, in which one SE round is assigned to each logical qubit within the element in turn before returning to the first, this strategy represents a uniform schedule across multiple logical qubits sharing the same ancilla resources. Dashed lines correspond to \textit{Priority SE}, which accounts only for the additional shuttling overhead of higher encoding ratios while ignoring the other co-hosted logical qubits entirely: syndrome extraction is performed exclusively for one logical qubit, with no intervening idling time between rounds. The relative improvement of Priority over Round-Robin SE at $d = 13$ is annotated for each encoding ratio.

As expected, increasing $m$ raises the LER per SE round, reflecting the cumulative noise overheads imposed on the physical qubits. At high encoding ratios ($m = 8$), the LER begins to plateau at large code distances, suggesting that the error correction capacity of the code is approaching saturation under the considered noise settings. Notably, this saturation occurs at higher code distances in the $T_{\phi}^\texttt{Store} = 800 \mu s$ case than in the $400 \mu s$ case, demonstrating that improving coherence times not only lowers the LER 
but also extends the range of encoding ratios that can be effectively supported before 
performance saturates.

The gap between Round-Robin and Priority SE schedules is also strongly noise-dependent. At 
$T_{\phi}^\texttt{Store} = 400 \mu s$, prioritizing the correction of a single logical qubit yields up to an $18\times$ improvement in LER, whereas at $T_{\phi}^\texttt{Store} = 800 \mu s$ the same configuration produces only a $6\times$ improvement. This disparity reflects the fact that longer storage-zone coherence times reduce the sensitivity to idling overhead: when $T_{\phi}^\texttt{Store}$ is large, qubits waiting in storage while other logical qubits are corrected decohere more slowly, and the cost of the Round-Robin schedule relative to Priority SE diminishes accordingly.

It should be noted that the Round-Robin and Priority schedules are inherently performing different tasks, and their direct comparison serves only to isolate the two sources of SE overhead (increased shuttling distance and increased idling time) from one another. Round-Robin SE simultaneously maintains the syndromes of all $m$ logical qubits, whereas Priority SE disregards the remaining $m - 1$ qubits, which will accumulate substantially higher LERs in the absence of correction. In light of this, all results in the subsequent sections use the Round-Robin schedule exclusively, with the understanding that the reported LERs represent a conservative bound that could be reduced further by more sophisticated SE prioritization strategies, that future compilation works may consider.

\subsection{Logical Qubit Scaling}

The memory benchmark under ancilla sharing was a necessary step to characterize the SE overhead introduced by higher encoding ratios, but improved memory performance was never the objective of this architectural proposal. The goal is to reduce the physical distance between logical qubits, thereby lowering the shuttling cost of transversal CNOTs and improving their LER. We now assess whether this benefit materializes in practice, and is worth the decrease in memory performance.

\begin{figure}
    \centering
    \includegraphics[width=\linewidth]{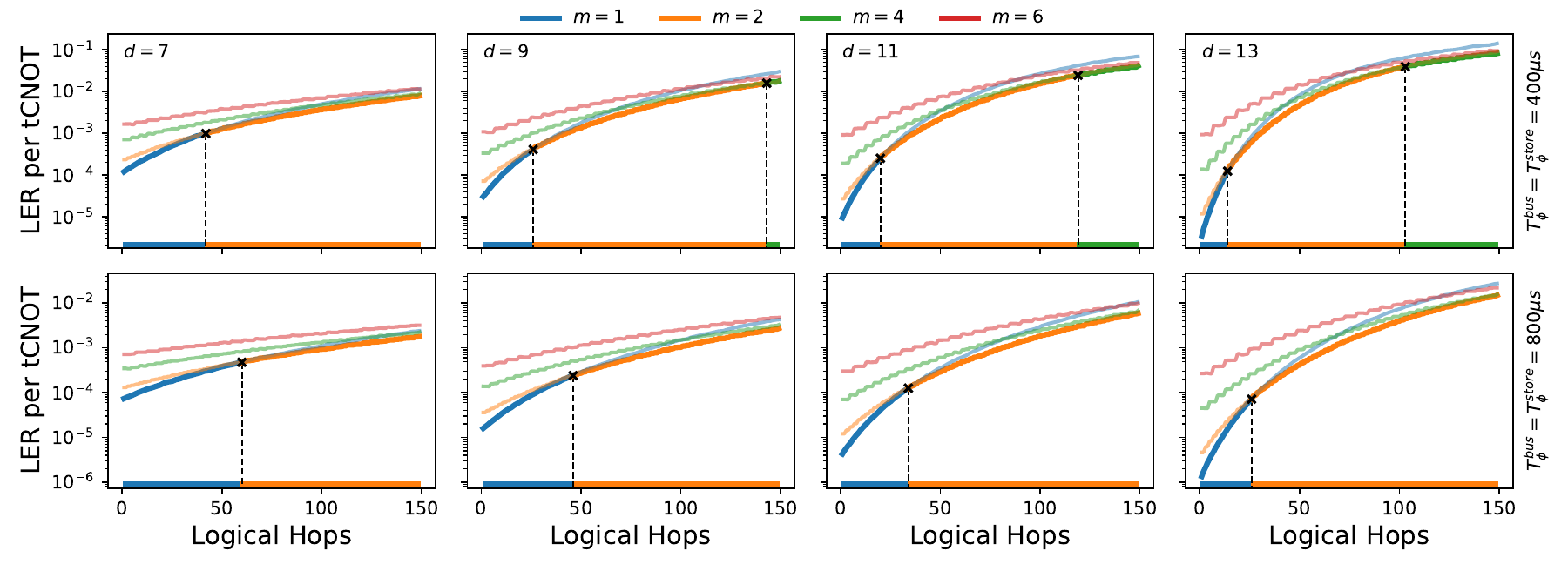}
    \caption{LER per tCNOT as a function of the number of logical hops, for code distances $d = 7$ to $d = 13$ and encoding ratios $m \in \{1, 2, 4, 6\}$. For $m > 1$, syndrome extraction follows the Round-Robin schedule. The best-performing encoding ratio at each logical hop count is highlighted on the corresponding curve and indicated in the horizontal bar at the bottom of each panel. The staircase profile of $m > 1$ curves reflects that logical qubits co-hosted within the same logical element can interact transversally without inter-element shuttling.}
    \label{fig:ratio_cnot}
\end{figure}

Figure~\ref{fig:ratio_cnot} reports the LER per tCNOT (each followed by one SE round) as a 
function of logical hops for code distances $d = 7$ to $d = 13$, under the same two noise 
configurations as the memory benchmark. Each curve corresponds to an encoding ratio $m \in 
\{1, 2, 4, 6\}$, and the best-performing encoding ratio at each logical hop count is 
highlighted both on the curve and in the horizontal bar at the bottom of each panel. The top 
row corresponds to $T_{\phi}^\texttt{Store} = 400 \mu s$ and the bottom row to 
$T_{\phi}^\texttt{Store} = 800 \mu s$. For all $m > 1$ cases, the Round-Robin SE schedule is used, providing a conservative lower bound on tCNOT performance under ancilla sharing.

The results show that higher encoding ratios are most beneficial at large code distances and large logical hop counts. For $T_{\phi}^\texttt{Store} = 400 \mu s$, an $m = 2$ encoding begins to outperform the $m = 1$ baseline at 14 logical hops, demonstrating that reduced inter-qubit distance can compensate for the SE overhead even at relatively modest separations. At short logical hop counts, however, $m > 1$ encodings perform considerably worse than the $m = 1$ baseline: while shuttling distances are always shorter for higher $m$, the idling overhead from Round-Robin SE significantly raises the LER. This penalty diminishes as the number of logical hops increases, until the connectivity advantage of ancilla sharing eventually takes over and the higher-$m$ encoding outperforms the lower-$m$ alternative.

A notable feature of the $m > 1$ curves is their staircase-like LER. When multiple logical qubits share the same ancilla set, they also share the operation zones used for transversal gates, meaning that logical qubits co-hosted within the same logical element can interact with one another without any inter-element shuttling. Consider, for example, the $m = 4$ logical element depicted in the bottom row of Figure~\ref{fig:ancilla_sharing}: transversal gates between any pair among logical qubits $A$, $B$, $C$, and $D$ are executed entirely within the element, incurring minimal shuttling overhead. The width of each horizontal step in the staircase corresponds precisely to the encoding ratio $m$, reflecting the number of logical qubits that can mutually interact without leaving their logical element.

Beyond the direct improvement in long-distance tCNOT LER, ancilla sharing carries significant practical benefits at the hardware level. A lower total quantum dot count reduces fabrication costs, shrinks the physical footprint of the processor and alleviates the cryogenic heat load~\cite{vandersypen2017interfacing, krinner2019engineering}. Perhaps most critically for near-term devices, fewer quantum dots reduce the probability of a manufacturing defect rendering the processor unusable: since semiconductor fabrication imperfections such as charge traps or lithographic variation can make individual dots inoperable~\cite{zwerver2022qubits}. Ancilla sharing thus improves scaling prospects by reducing the number of quantum dots that must all function correctly simultaneously.

\section{Shuttling Through Junctions}
\label{sec:junctions}

The one-dimensional shuttling bus adopted throughout the preceding sections was a deliberate simplification, chosen to isolate the core architectural trade-offs in a controlled setting. We now relax this constraint and consider how extending the architecture to a 2D grid of shuttling tracks can drastically reduce the distance between logical qubits \cite{yenilen2025performance, chadwick2025manufacturable}, at the cost of only a marginal increase in shuttling overhead.

\begin{figure}
    \centering
    \includegraphics[width=0.75\linewidth]{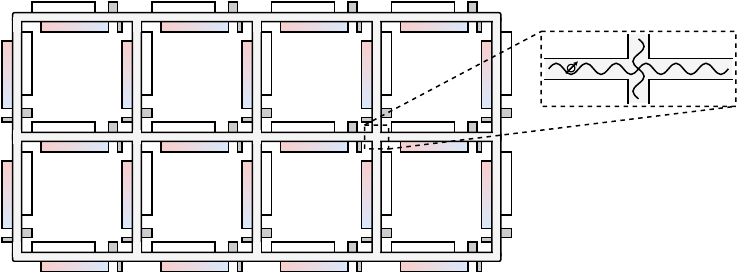}
    \caption{Two-dimensional shuttling bus architecture formed by introducing perpendicular shuttling tracks connected to the horizontal bus at regular intervals. Each intersection between a horizontal and a vertical track is implemented as a single additional quantum dot that allows a spin to either continue along its current direction or change direction, incurring a synchronization overhead when a direction change occurs. Logical elements are arranged on the resulting 2D grid, reducing the worst-case inter-qubit shuttling distance from $N - 1$ logical hops in the 1D case to $2\sqrt{N}$ logical hops in the 2D case for an architecture of $N$ logical qubits.}
    \label{fig:2d_connectivity}
\end{figure}

Rather than confining all logical elements to a single linear track, we introduce perpendicular shuttling lanes connected to the horizontal bus at regular intervals, forming a 2D grid of logical elements. The intersection between a horizontal and a vertical track requires only a single additional quantum dot (a \textit{junction}) shared between both tracks, which allows a spin to either continue along its current direction or change direction with minimal overhead. Preliminary experimental results suggest that direction changes at a junction incur only a short synchronization penalty (assumed here to be $50\,\text{ns}$, an arbitrary number that depends on the lab equipment used), while straight-through crossings introduce no additional time cost. This extra quantum dot slightly enlarges each logical element, adding a small but bounded shuttling overhead. Figure~\ref{fig:2d_connectivity} illustrates the resulting 2D architecture, with one junction placed between each pair of neighboring logical elements. All noise parameters are fixed to the previously discussed $T_{\phi}^\texttt{Store} = 800 \mu s$ configuration.

Given the simplifying assumptions underlying the junction model, the results in this section should be interpreted as a promising avenue for further improving the connectivity of the proposed architecture, rather than as a precise quantitative prediction of its performance. A more detailed experimental and numerical characterization of junction behavior is left for future work.

Importantly, the memory capabilities of each logical qubit are unaffected by the introduction of junctions, as SE continues to operate entirely within the domain of a single logical element, for which the layout has already been optimized.

To characterize the tCNOT LER in the 2D setting, we move away from the logical-hop benchmarking used for the 1D architecture, as a direct hop-count comparison between 1D and 2D architectures would be misleading, as the same number of logical hops corresponds to fundamentally different physical distances and connectivity structures in the two cases. Instead, we construct a full architecture containing $N$ logical qubits and benchmark the tCNOT between the two most distant logical qubits, for both architectures simultaneously. In the 1D case, this worst-case gate involves two logical qubits separated by $N-1$ logical hops. In the 2D case, the improved connectivity reduces the worst-case separation to $2\sqrt{N}$ logical hops, with two direction changes along the shuttling path, yielding a substantially shorter route for all but the smallest architectures.

\begin{figure}
    \centering
    \includegraphics[width=\linewidth]{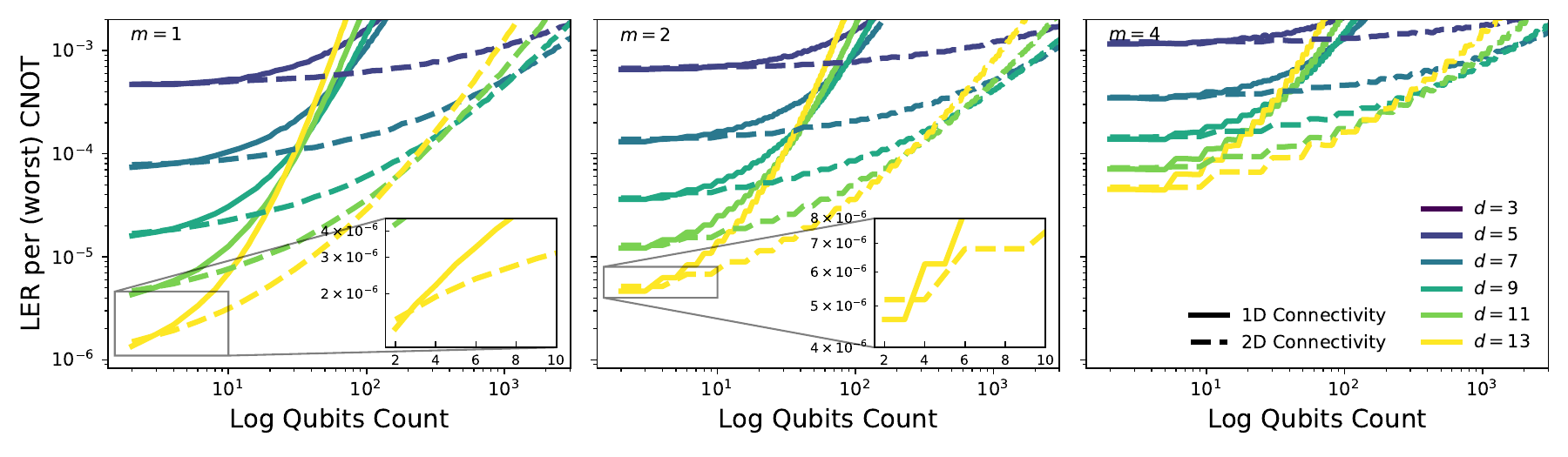}
    \caption{LER per tCNOT for the two most distant logical qubits in an architecture of $N$ logical qubits, shown for $N$ up to $300$ and encoding ratios $m \in \{1, 2, 4\}$. Results are reported for both 1D (linear bus) and 2D (grid) connectivity layouts.}
    \label{fig:increasing_log_count}
\end{figure}

Figure~\ref{fig:increasing_log_count} reports the tCNOT LER for architectures ranging from $N = 2$ to $N = 300$ logical qubits, for encoding ratios $m \in \{1, 2, 4\}$ distributed across both 1D and 2D connectivity layouts, across all code distances considered. The advantage of 2D connectivity appears even for modest qubit counts, the 2D architecture consistently achieves lower tCNOT LERs than its 1D counterpart across all encoding ratios and code distances tested.

From the inset axes that zoom in into the architectures with less than 10 logical qubits, we can see that 1D approaches outperform 2D one whenever logical qubits are neighbouring. This is important to notice since, the shown results benchmark the tCNOT across the two more distant logical qubits in the architecture. However, during the execution of an algorithm, tCNOTs will happen across a variety of distances between qubits, for which it is important to notice the improved performance of 1D arrays.

The inset axes zoom into the small-$N$ regime and reveal an important observation. When the two logical qubits involved in a tCNOT are immediate neighbors, the 1D layout slightly outperforms the 2D one: in this case, the junction introduces a small but non-zero overhead without providing any reduction in shuttling distance, since the qubits are already as close as possible. This is worth keeping in mind when interpreting the main figures, which report the LER of the \textit{worst-case} tCNOT in an architecture of $N$ logical qubits, a metric that naturally favors the 2D layout as $N$ grows. In practice, a quantum algorithm will issue transversal gates across the full spectrum of inter-qubit distances, and for short-range interactions the 1D architecture remains competitive. The relative advantage of each layout will therefore depend on the specific logical circuit being executed and the distribution of gate distances it entails.

\section{Application to Magic State Distillation}
\label{sec:distillation}

Extending quantum computation beyond the Clifford group to achieve universal fault-tolerant operation requires the efficient production of high-fidelity non-Clifford resource states \cite{eastin2009restrictions}. This is typically accomplished through Magic State Distillation (MSD), in which many copies of low-fidelity magic states are distilled into fewer, higher-fidelity ones using only Clifford operations and measurements, via a dedicated Magic State Factory (MSF)~\cite{bravyi2012magic, ding2018magic, litinski2019magic}. In this section, we assess how the transversal gate capabilities of the proposed architecture can be leveraged to implement such a distillation routine efficiently.

\begin{figure}
    \centering
    \includegraphics[width=\linewidth]{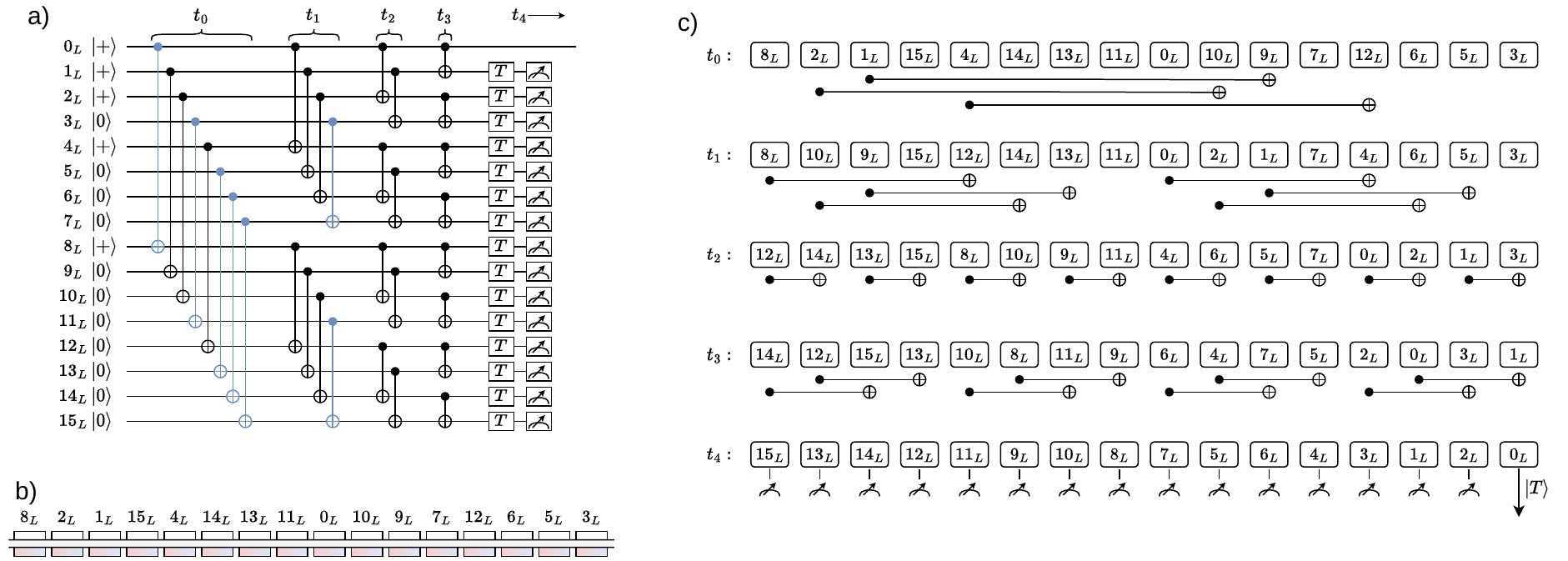}
    \caption{\textbf{a)} \textit{15-to-1} $T$-state distillation circuit. At the end of the circuit, 15 qubits are measured and the remaining unmeasured qubit holds the distilled high-fidelity $T$ state. \textbf{b)} Optimal 1D logical qubit placement for the distillation circuit, obtained by applying the Quantum Reverse Mapping methodology. (\textit{c}) Evolution of the logical qubit layout across the four CNOT time steps $t_0$--$t_3$, showing how symmetric shuttling progressively exchanges qubit positions.}
    \label{fig:distillation}
\end{figure}

We consider the \textit{15-to-1} $T$-state distillation circuit of~\cite{beverland2021cost}, depicted in Figure~\ref{fig:distillation} a). The protocol uses 16 logical qubits, 15 of which are measured at the end of the circuit, and the remaining unmeasured qubit holds the distilled high-fidelity $T$ state, ready to be injected whenever a $T$ gate is required during computation. Note that the gates highlighted in blue are redundant and can be omitted without affecting the output. The circuit decomposes into four time steps of parallel CNOTs, followed by the injection of noisy $T$ gates into qubits $1_L$ through $15_L$ and their subsequent measurement.

To maximize the performance of this distillation routine on the proposed architecture, we apply the Quantum Reverse Mapping methodology of~\cite{escofet2026synthesizing} at the \textit{logical} level, optimizing the placement of the 16 logical qubits along the 1D shuttling bus so that the overall tCNOT cost of the circuit is minimized. This mirrors the approach taken at the physical level in~\cite{escofet2026synthesizing}, where the same methodology was used to optimize the arrangement of physical qubits for syndrome extraction, and demonstrates the versatility and cross-layer applicability of the Quantum Reverse Mapping framework. The MILP formulation is updated to enforce that, whenever two logical qubits interact via a tCNOT, they exchange positions in the 1D bus, thus implementing the symmetric shuttling strategy shown to minimize tCNOT LER in previous results (see Figure~\ref{fig:transv_cnot_sym}). The resulting optimal logical placement is shown in Figure~\ref{fig:distillation} b).

Figure~\ref{fig:distillation} c) traces the evolution of the logical qubit layout across the four CNOT-filled time steps, illustrating how the arrangement changes as symmetric shuttling progressively exchanges qubit positions. A particularly convenient feature of the optimized layout is that the unmeasured qubit (the one holding the distilled $T$ state) ends up at one side of the architecture after the circuit completes. This placement is advantageous for integration into a larger processing unit. By connecting this edge of the MSF to the main 2D shuttling bus via a junction, the distilled state can be injected into the computation with minimal additional shuttling.

From the optimized layout, we can assess the tCNOT requirements of the full distillation circuit: three 8-hop tCNOTs in time step $t_0$, six 4-hop tCNOTs in $t_1$, eight 1-hop tCNOTs in $t_2$, and eight 2-hop tCNOTs in $t_3$. We estimate the overall logical failure probability of the distillation subroutine by combining the previously characterized per-operation LERs (for each $n$-hop tCNOT and for the memory baseline during idle time steps) via the standard additive model~\cite{litinski2019game, beverland2021cost, filippov2025architecting}:
\begin{equation}
    \text{LER}_{\texttt{circ}} = 1 - \prod_{i=1}^{N} \left(1 - \text{LER}_i\right),
\end{equation}
where $\text{LER}_i$ is the measured error rate of the $i$-th constituent operation. This model allows us to assess how architectural choices (such as code distance, encoding ratio, and shuttling policy) affect the overall distillation LER, without requiring a full circuit-level simulation that would also need to account for $T$-gate injection noise. The results here should therefore be understood as an assessment of how well the architecture can be tuned to minimize the tCNOT contribution to the overall distillation error, rather than as an absolute prediction of end-to-end distillation fidelity.

\begin{figure}
    \centering
    \includegraphics[width=\linewidth]{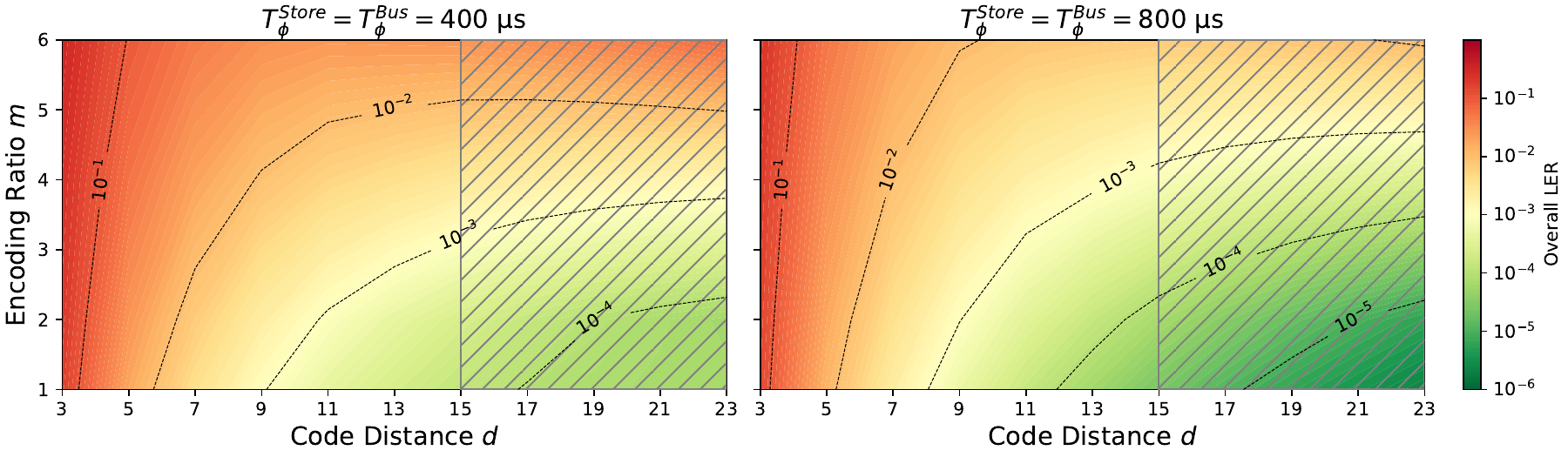}
    \caption{Estimated logical failure probability of the \textit{15-to-1} $T$-state distillation circuit as a function of code distance $d$ and encoding ratio $m$.}
    \label{fig:ler_distillation}
\end{figure}

Figure~\ref{fig:ler_distillation} reports the overall circuit LER as a function of code 
distance and encoding ratio, under the $T_{\phi}^\texttt{Store} = 400 \mu s$ and $T_{\phi}^\texttt{Store} = 800 \mu s$ noise configurations. Results for code distances $d \geq 15$ are projected from lower-distance  benchmarks using an exponential fit (shaded region).

Two observations stand out. First, owing to the small number of logical qubits in the MSF (16) and the compactness of the optimized layout, the longest tCNOT in the circuit spans only 8 logical hops. As a result, higher encoding ratios offer no benefit for this particular routine: the inter-qubit distances are already short enough that the SE overhead of ancilla sharing outweighs any connectivity gain. This will not be the case for practical algorithms involving hundreds or thousands of logical qubits, where long-range interactions are unavoidable and higher encoding ratios are expected to be advantageous. Second, while the impact of increasing $T_{\phi}^\texttt{Store}$ from $400$ to $800\,\mu\text{s}$ appeared modest at the level of individual operations, the cumulative effect across all memory and tCNOT contributions to the distillation circuit yields a substantially larger difference in overall LER, underscoring the importance of coherence time improvements at the system level.

Taken together, the results of this section demonstrate the power of applying consistent optimization principles across multiple levels of the quantum computing stack. At the physical level, Quantum Reverse Mapping minimizes the overhead of syndrome extraction by optimally placing physical qubits within each logical element. At the logical level, the same methodology minimizes the tCNOT cost of the distillation circuit by optimally arranging logical qubits along the shuttling bus and enforcing symmetric shuttling. The combination of these two co-design steps yields an architecture in which both the memory and computation layers are simultaneously tuned for performance, providing a concrete blueprint for how spin-qubit shuttling architectures can be systematically optimized from the physical qubit to the 
algorithmic primitive.

\section{QEC Code Selection}
\label{sec:discussion}

Although this work uses the surface code as a concrete substrate, the architectural principles and design methodology introduced here are broadly applicable to any QEC code.

A well-known limitation of the surface code is its unfavorable $n/k$ ratio: encoding $k$ logical qubits requires $n = d^2$ physical qubits per logical qubit, leading to long logical elements and, consequently, large shuttling distances for transversal two-qubit interactions. Replacing the surface code with a QEC code that offers a better $n/k$ ratio would directly shorten the logical elements, reduce inter-qubit shuttling distances, and lower the shuttling-induced dephasing error per tCNOT, achieving the same connectivity improvement as a higher encoding ratio $m$, but without the SE overhead that ancilla sharing entails.

\begin{figure}
    \centering
    \includegraphics[width=0.5714\linewidth]{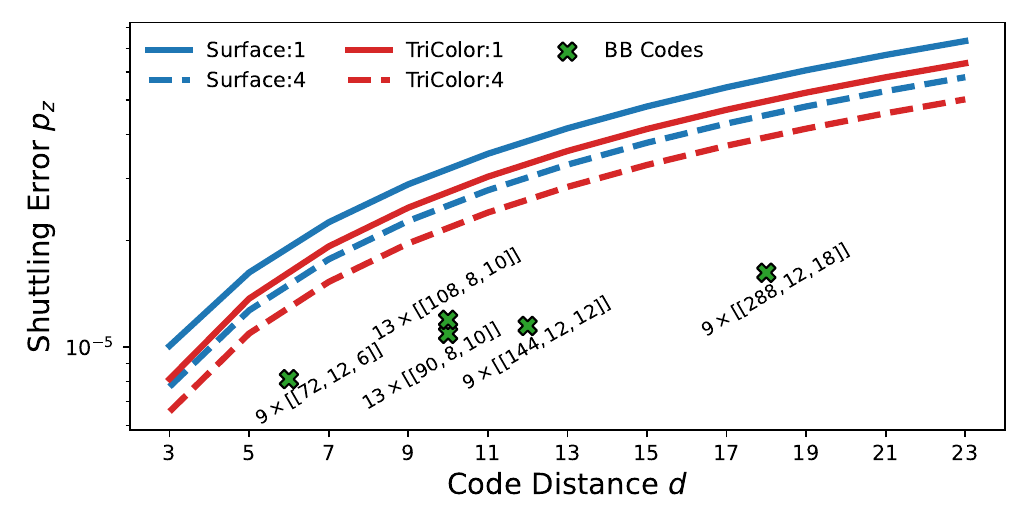}
    \caption{Pauli-$Z$ error accumulated by each data qubit when traversing 100 logical elements for a transversal CNOT, shown for the surface code ($m = 1$ and $m = 4$), triangular color codes ($m = 1$ and $m = 4$), and members of the bivariate bicycle (BB) code. Noise parameters follow the settings of Table~\ref{tab:parameters}, with $T_{\phi}^\texttt{Store} = T_{\phi}^\texttt{Bus} = 800 \mu s$.}
    \label{fig:other_codes_shuttling}
\end{figure}

A full LER characterization across different QEC codes is beyond the scope of this work; however, to provide intuition on the expected gains, Figure~\ref{fig:other_codes_shuttling} reports the Pauli-$Z$ error accumulated by each data qubit when traversing 100 logical elements to reach its destination for a transversal CNOT. Curves are shown for the surface code at encoding ratios $m = 1$ and $m = 4$, for triangular color codes~\cite{bombin2006topological, landahl2011fault} at $m = 1$ and $m = 4$, and as discrete points for five members of the bivariate bicycle (BB) code family~\cite{bravyi2024high}.

The results confirm the expected trend: codes with a better $n/k$ ratio accumulate substantially less shuttling-induced error per logical hop, with BB codes offering the most favorable scaling among the families considered. Importantly, the practical hardware benefits of higher encoding ratios discussed in the previous section (reduced manufacturing cost, smaller device footprint, lower cryogenic power consumption, and improved fabrication yield) apply equally when switching to a more qubit-efficient QEC code, since both strategies reduce the total number of physical quantum dots required per logical qubit.

\section{Conclusions}
\label{sec:conclusions}

In this work, we have presented a multi-qubit shuttling bus architecture for spin-qubit processors that supports both reliable logical qubit storage and transversal two-qubit gates, building directly upon the single-logical-qubit design optimized for syndrome extraction in~\cite{escofet2026synthesizing}. By concatenating logical elements along a shared shuttling bus and leveraging the inherent long-range connectivity of spin-qubit shuttling, the proposed architecture achieves all-to-all logical connectivity while keeping the overhead of each QEC cycle confined to the domain of a single logical element. We characterized the architecture through a realistic noise model and benchmarked its performance across both memory and transversal CNOT experiments over a wide range of noise parameters representative of current and near-future spin-qubit devices.

We showed that symmetric shuttling consistently outperforms asymmetric shuttling for transversal gates, reducing the LER by up to $60\%$ at large code distances. We further demonstrated that ancilla sharing across multiple logical qubits within a single logical element reduces the physical distance between logical qubits and improves the LER of long-range transversal gates (at the cost of increased SE latency), while simultaneously offering practical hardware benefits. Extending the architecture from a 1D bus to a 2D grid of shuttling tracks via junctions further reduces the worst-case inter-qubit distance from $O(N)$ to $O(\sqrt{N})$ logical hops, yielding rapid and consistent improvements in tCNOT LER across all encoding ratios and code distances tested.

We applied the Quantum Reverse Mapping methodology at the logical level to optimize the placement of logical qubits for a \textit{15-to-1} magic state distillation circuit, demonstrating that the same co-design principles that minimize syndrome extraction overhead at the physical level can be used to minimize tCNOT cost at the logical level. The resulting optimized layout naturally places the distilled $T$ state at the boundary of the factory, facilitating its injection into a larger processing unit with minimal additional shuttling. This cross-layer application of the methodology showcases its versatility demonstrating that jointly optimizing layers of the stack produces greater overall gains than optimizing each layer independently.

This work illustrates the tight and consequential interplay between architectural design and the QEC layer of the quantum computing stack. The choice of code, its $n/k$ ratio, its stabilizer weight, and its syndrome-extraction schedule all directly shape the physical layout of the processor, the shuttling overhead of logical operations, and ultimately the achievable logical error rates. Navigating this co-design space systematically, rather than treating the hardware and the code as independent choices, is the central methodological contribution of this work, and provides a principled and reusable blueprint for optimizing quantum computing architectures as they progress toward fault-tolerant operations.

The ultimate goal of architecture design is not to optimize isolated primitives, but to maximize the end-to-end performance of the quantum routines that are expected to deliver computational advantage. As quantum processors scale toward the regime where known advantageous algorithms become executable, the architecture design choices explored here (QEC code selection, encoding ratio, shuttling policy, bus topology, and cross-layer placement optimization) will collectively determine whether the available hardware can sustain the logical fidelity that quantum advantage demands.

\bibliography{sn-bibliography}
\end{document}